# Simultaneously imaging of dielectric properties and topography in a $PbTiO_3$ crystal by near-field scanning microwave microscopy

Y.G. Wang[a], M. E. Reeves[a,b], F.J. Rachford[b]
[a] *The George Washington University, Washington, DC 20052*
[b] *Naval Research Laboratory, Washington, DC 20375*

We use a near-field scanning microwave microscope to simultaneously image the dielectric constant, loss tangent, and topography in a $PbTiO_3$ crystal. By this method, we study the effects of the local dielectric constant and loss tangent in the geometry of periodic domains on the measured resonant frequency, and quality factor. We also carry out theoretical calculations and the results agree well with the experimental data and reveal the anisotropic nature of dielectric constant.

Ferroelectrics have a nonlinear dielectric constant which can be tuned by a bias electric field. This property makes ferroelectrics promising candidates for microwave devices such as tunable phase shifters and filters.[1] In order to improve the material properties for these and similar tunable microwave devices, it is essential to measure the microwave dielectric properties accurately and to map local property variations. So far, several types of near-field scanning microwave microscopes have been developed, each with submicron resolution: the $\lambda/4$,[2] coaxial line,[3] and ring resonators.[4] However, most observations have been made in a soft-contact mode where the contribution of the topography is inseparable from the dielectric properties, and tip damage inevitable. Thus, it is desirable to develop a means to simultaneously obtain microwave dielectric properties and topography. In addition, quantitatively characterizing the microwave properties on a microscopic scale remains a challenge, because the Coulomb force between the tip and sample depends not only on the material, but also on the shape and size of the regions with properties different from the host material.

Anisotropy is often an essential property of dielectrics, whether it arises from intrinsic[5] or extrinsic sources. For example, stress-induced anisotropy in thin-film strontium barium titanate[6] is the probable source of discrepancies between measurements of permittivity made by the techniques of interdigitated electrodes and scanning microwave microscopy. Electrodynamic calculations show that the dominant component of the electrical field is in-plane for interdigitated electrodes but out-of-plane for scanning microwave microscopy. Thus the anisotropic nature of the dielectric matrix should be taken into account when quantitatively characterizing material properties. In this paper we show that the anisotropy of the dielectric constant can be resolved by careful comparison between calculations and experimental data.

This paper reports measurements on a $PbTiO_3$ crystal where the dielectric constant, loss tangent and topography are obtained simultaneously. $PbTiO_3$ is a good candidate for microwave applications after doping,[7] hence its microwave dielectric properties have been well-characterized.[8] At room temperature, it has parallel $a$ and $c$ domains[9,10] which give characteristic signatures in x-y scans of resonant frequency ($f_0$), quality factor (Q), and surface topography. In our experimental configuration, the $a-c$ domain walls are the (101) planes as illustrated schematically in Fig. 1.

The near-field scanning microwave microscope used in our study consists of a 1.75 GHz, $\lambda/4$ coaxial resonator[2] which is driven by an HP 8753D network analyzer. A polished tungsten STM tip protrudes from the central conductor of the cavity and provides close coupling to the sample under study. Thus, the dielectric constant and loss tangent can be calculated from the measured resonant frequency and quality factor, and near-field microscopy in the submicron range can be realized. To accurately control the placement of the sample in three dimensions, piezoelectric actuators are used for positioning.

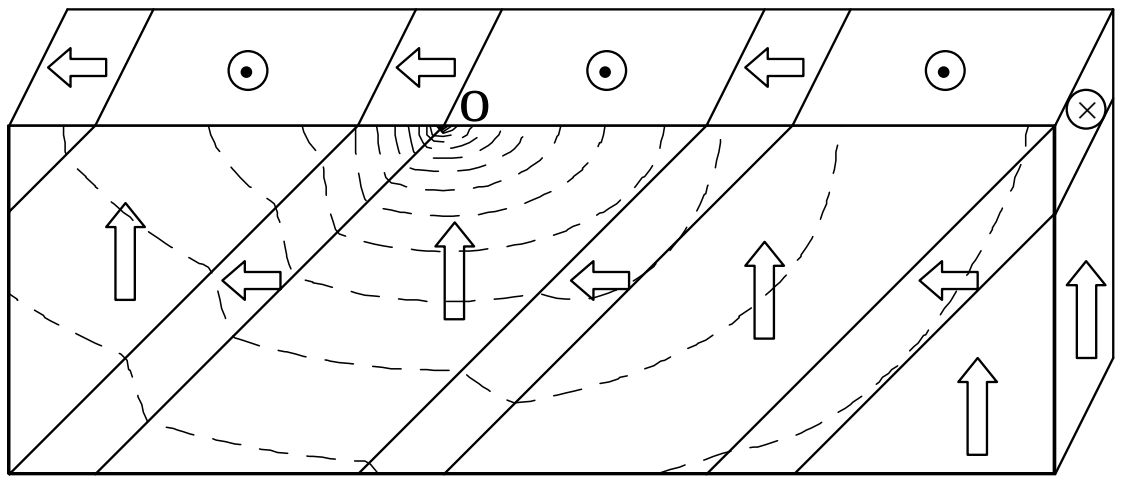

FIG. 1. A sketch of the $a-c$ domain structures in a $PbTiO_3$ crystal. The arrows show the directions of the spontaneous polarization, and the dotted lines are the equipotentials.

We compare our data to finite-element calculations of the tip-sample response. A static approximation is used considering that the tip size and the tip-sample distance are much smaller than the wavelength (17 cm at 1.75 GHz). We model the tip as a cylinder capped by a cone with a spherical end, all held at a constant potential. We also use a commercial software package[11] to calculate the field distribution by variable-mesh finite-element analysis. The changes of resonant frequency and quality factor are then obtained using perturbation theory. That is, we assume that the small contributions of the fields generated within the sample are capacitively coupled to the cavity.

Figure 2 shows the dependence of $f_0$ on the tip-sample distance, h, above an $a$ and a $c$ domain in a $PbTiO_3$

crystal. As the tip approaches the sample (increasing displacement in Fig. 2), $f_0$ first increases rapidly, then the slope drops to zero upon touching. Thus we can control the tip height by measuring $df_0/dh$ while adjusting the sample's vertical position. The resulting resonant frequency, quality factor, and topography are shown in Fig. 3. These exhibit the expected periodic structure of alternating $a$ domains ($\epsilon_a$=105, $\tan\delta_a$=0.04) and $c$ domains ($\epsilon_c$=35 and $\tan\delta_c$=0.08). The measured bending angle, $\theta$, at the $a-c$ domain wall is $3.8 \pm 0.2°$ in agreement with results from atomic force microscopy[9] and the theoretical value of 3.65°[10]. In the Q image of Figure 3(b) there are some holes with various shapes and sizes, such as the one denoted by F. These are due to internal defects originating from the flux-growth process.

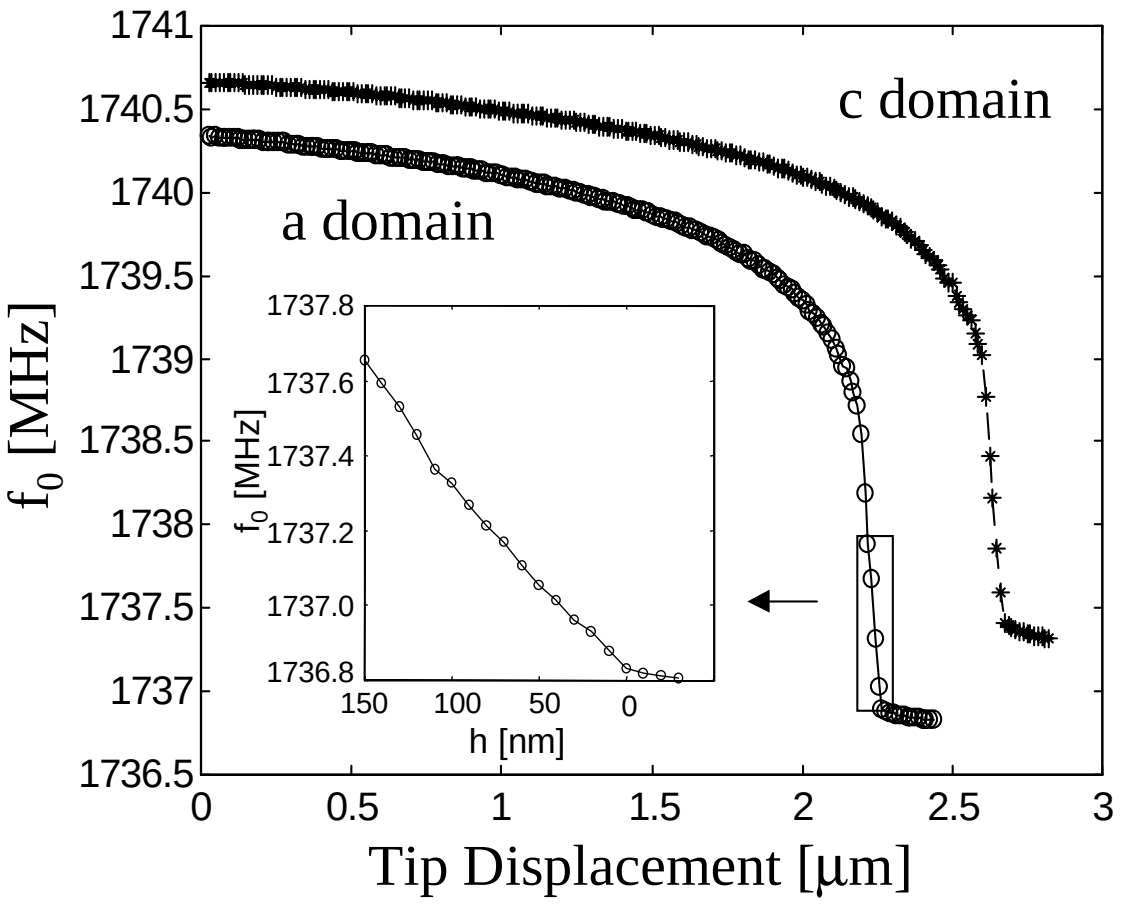

FIG. 2. Change of resonant frequency, $f_0$, with tip height for $a$ (circles) and $c$ (stars) domains in a $PbTiO_3$ crystal. The inset shows the dependence of $f_0$ on tip-sample distance h close to the sample.

As seen from the cross-section profiles (Fig. 4), the maximum and minimum positions of Q correspond to the centers of the $a$ and $c$ domains, respectively. In contrast, the profile of $f_0$ is distinctly different with its maximum and minimum positions close to the $a-c$ domain walls. The apparently peculiar shape of the $f_0$ profile originates from the anisotropy of the dielectric constant and the tilt of the $a-c$ domain walls with respect to the surface normal.

To verify these results, we have calculated the $f_0$ profile under the following three conditions: (i) the domains are isotropic with walls perpendicular to the surface normal; (ii) the crystal is composed of tilted, isotropic slabs with alternatively changing dielectric constants, 105 and 35; (iii) the more realistic model consisting of alternating tilted, anisotropic domains. The third case models the data well as seen by comparing the solid curve in Fig. 5(a) with the cross-section profile in Fig. 4(b). We attribute this result to the fact that though the electrical field is primarily perpendicular to the surface, there is a significant parallel component, which probes the anisotropy. We also calculate the Q profile under the circumstance (iii); the results shown in Fig. 5(b), agree with the experimental data.

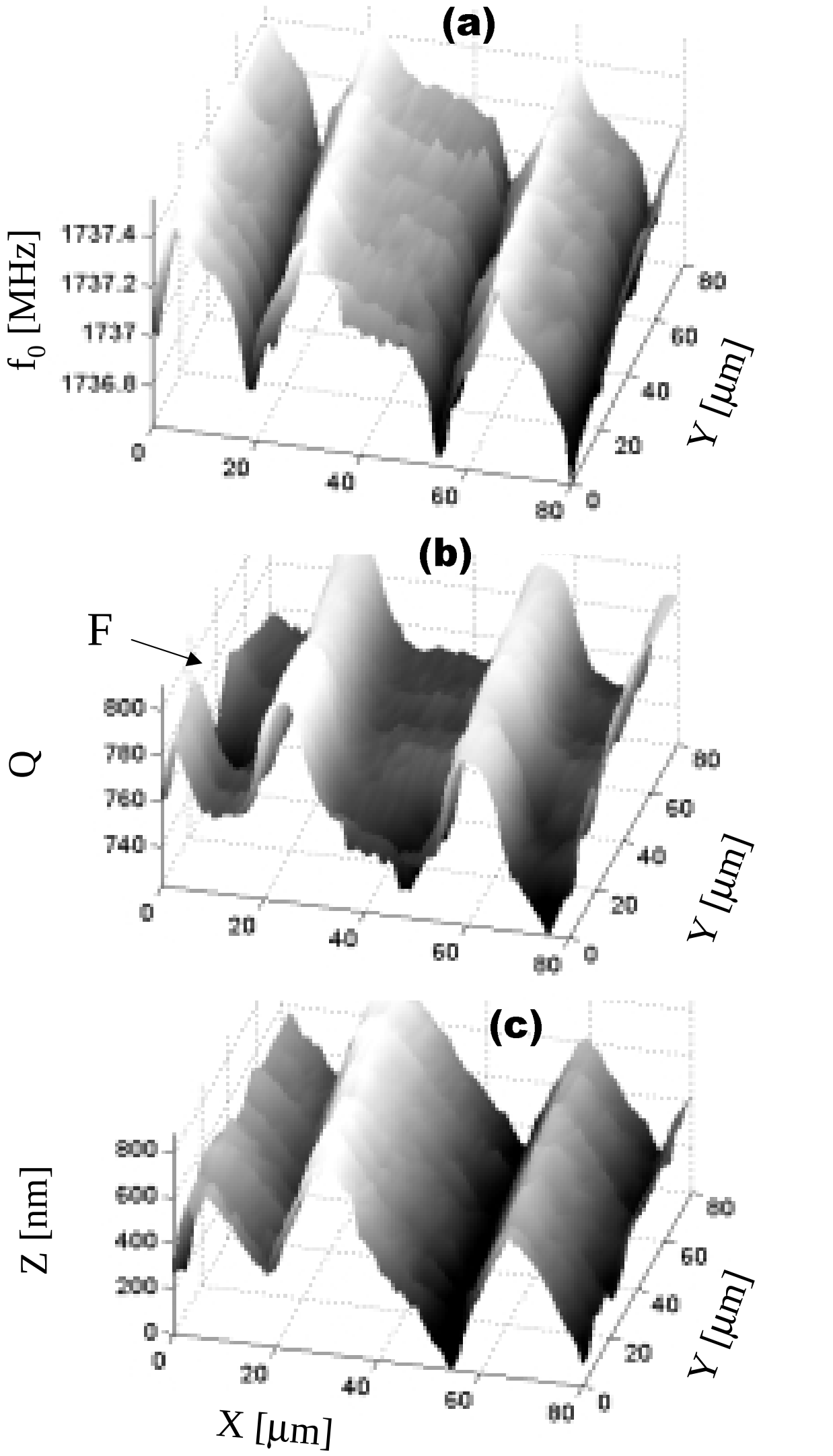

FIG. 3. Near-field images of a $PbTiO_3$ crystal showing simultaneously obtained (a) resonant frequency, (b) quality factor, and (c) topography. The $\theta$ denotes the bending angle of the surface at the $a-c$ domain wall.

Our findings show the importance of combining near-field microwave imaging with topographic information and with precise calculations of the electric fields used to probe the dielectric properties. From this approach, a determination can be made of the magnitude and anisotropy of $\epsilon$ and $tan\delta$. This is particularly significant when comparing measurements made by different techniques on barium strontium titanate films: an important material for microwave applications which has significant in-plane vs. out-of-plane anisotropy.

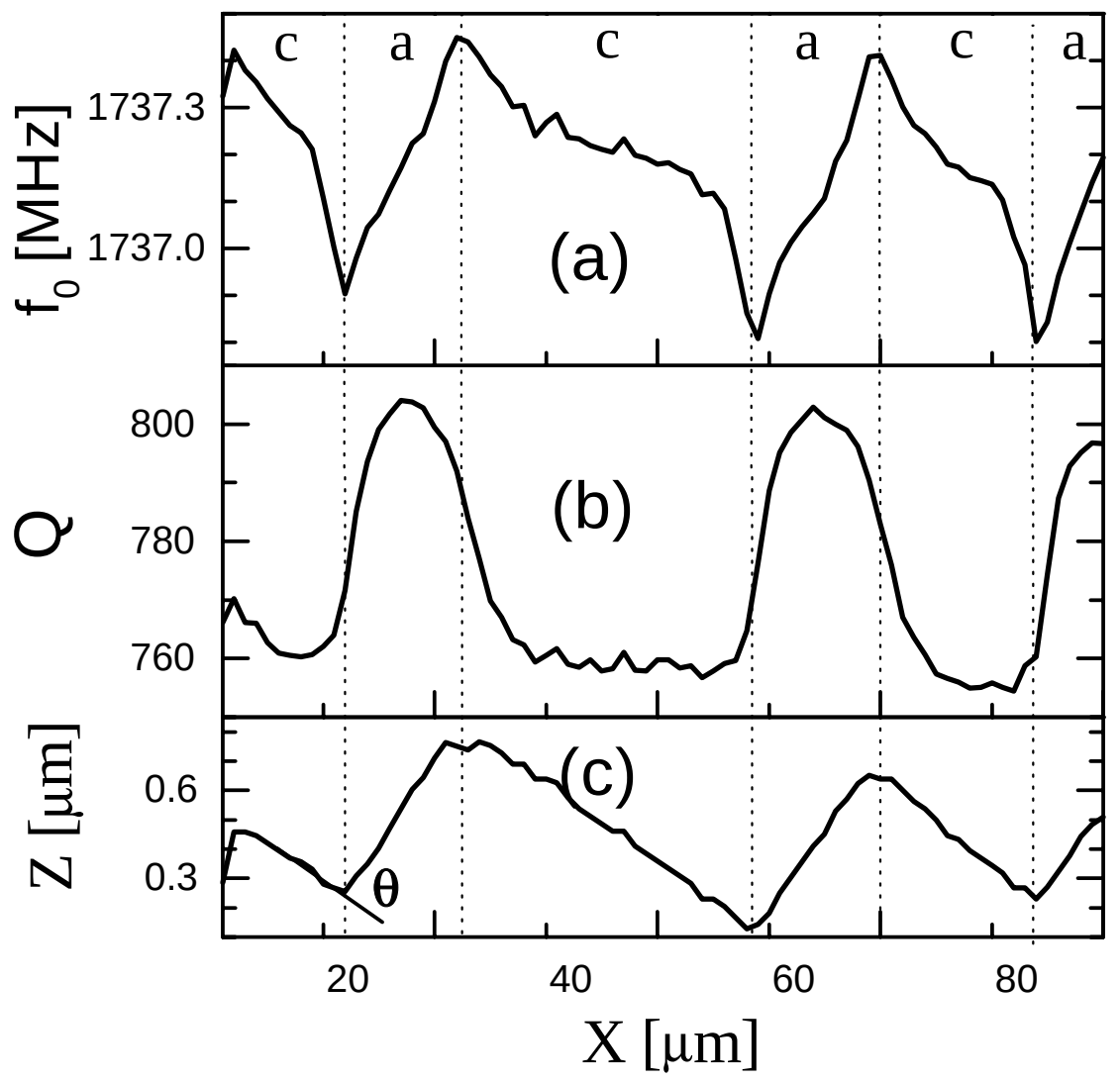

FIG. 4. The measured cross-section profile at y=70$\mu$m for (a) resonant frequency, (b) quality factor, and (c) topography. The angle $\theta$ denotes the bending angle of the surface at the $a-c$ domain wall.

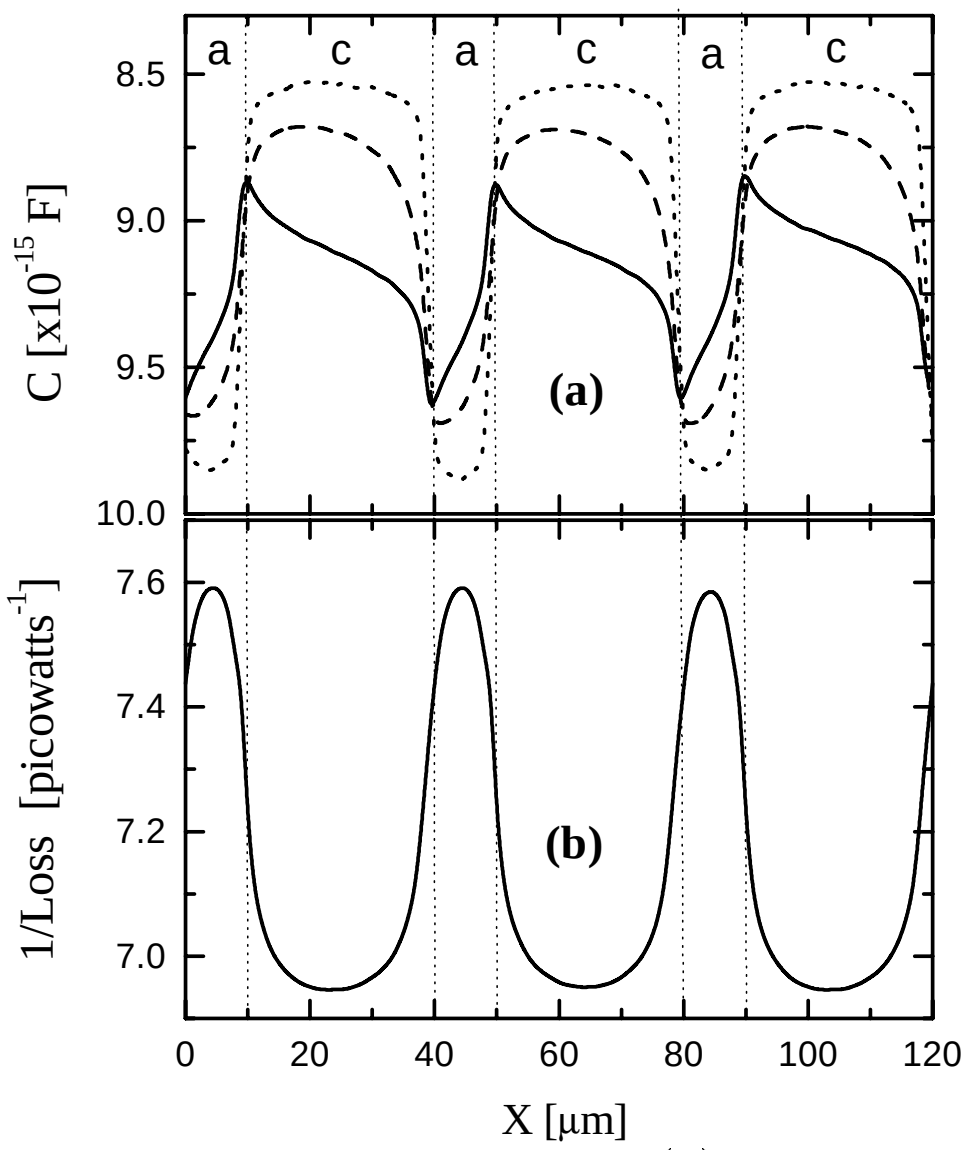

FIG. 5. The calculated profiles of (a) tip-sample capacitance and (b) loss for a real $PbTiO_3$ crystal (solid line) with the domain configuration shown in Fig. 1, for an imaginary crystal where the dielectric constants are isotropic (dashed line) and for an imaginary crystal where the $a-c$ domain wall is normal to the surface. The dotted lines separate areas of $a$ and $c$ domains.

Additional data which underscores the importance of modeling come from measurements of $PbTiO_3$ $a$ and $c$ domains with different widths. The images obtained indicate that the contrast in $f_0$ and Q increases with domain width, so that one should always take the impurity size into account when transforming the measured $f_0$ and Q into permittivity and loss tangent. To further explore this, we have simulated the shift of $f_0$ after introducing cylindrical impurities with various permittivities and sizes into a uniform dielectric ($\epsilon$=100). The calculated shift increases nearly logarithmically with impurity size, but tends to saturate when the impurity size is ten times larger than the tip radius of 1 $\mu$m. For an impurity with $\epsilon$=120, the contrast goes below a detection limit of $10^{-5}$ when the impurity radius is one order of magnitude smaller than the tip radius. The shift of $f_0$ increases with contrast in permittivity, as expected.

Our recent study on a barium strontium titanate film[12] has revealed the importance of simultaneously imaging the near-field microwave properties and the topography. We found ring-shaped regions with higher resonant frequency and quality factor than the surroundings in the film. The rings are 5$\sim$10$\mu$m in diameter, and about 100nm higher than the surroundings as determined by the simultaneously obtained topography . The topography is further confirmed by atomic force microscopy and provides one possible explaination for the observed inhomogeneities in the resonant frequency and the loss tangent.

Acknowledgements: The work is supported by DARPA. The authors would like to thank Mr. Z. Wang of the Institute of Crystal Materials of Shandong University for supplying the crystal and Drs. J.M. Byers and M. M. Miller for helpful discussions.

---